\documentclass[nofootinbib,floats,letter,floatfix,prl,aps,10pt]{revtex4}

\usepackage{dcolumn,epsfig,minitoc}
\usepackage{hyperref}
\usepackage{amssymb,amsmath}
\usepackage[usenames]{color}

\def\b{\beta}\def\d{\delta}
\def\f{\phi}\def\g{\gamma}
\def\l{\lambda}\def\m{\mu}\def\n{\nu}\def\r{\rho}
\def\y{\eta}\def\x{\xi}
\def\ee{\varepsilon}

\def\O{\Omega}
\def\OM{dx_{i}dx^{i}¥}\def\S{\Sigma}
\def\l{\lambda}

\def\inf{\infty}

\def\ex{{\rm e}}

\def\sch{Schwarzschild }

\def\o1{$O_{1}$}
\def\o2{$O_{2}$}

\def\ee{e^{2\y}}\def\en{e^{2\n}}\def\er{e^{2\r}}\def\eax{e^{2a\x}}

\def\lb{\label}
\newcommand{\beq}{\begin{equation}}
\newcommand{\eeq}{\end{equation}}
\newcommand{\bea}{\begin{eqnarray}}
\newcommand{\eea}{\end{eqnarray}}

\begin{document}
\title{ Black brane solutions and their  solitonic extremal limit 
in Einstein-scalar gravity }
\author{Mariano Cadoni}
\affiliation{Dipartimento di Fisica, Universit\`a di Cagliari and INFN, Sezione di
Cagliari - Cittadella Universitaria, 09042 Monserrato, Italy. }
\author{Salvatore Mignemi}
\affiliation{Dipartimento di Matematica, Universit\`a di Cagliari and INFN, Sezione di
Cagliari - viale Merello 92, 09123 Cagliari, Italy. }
\author{Matteo Serra}
\affiliation{Dipartimento di Fisica, Universit\`a di Cagliari and INFN, Sezione di
Cagliari - Cittadella Universitaria, 09042 Monserrato, Italy. }
\date{\today}


\date{\today}

\begin{abstract}
We  investigate static, planar, solutions of Einstein-scalar gravity admitting 
an anti-de Sitter (AdS) vacuum. When the squared mass  of the scalar field is positive and the 
scalar potential can be derived from a superpotential, minimum energy 
theorems indicate the existence of  a   scalar soliton. On the other 
hand, for these models, no-hair theorems forbid the existence of 
hairy 
black brane solutions 
with AdS asymptotics. By considering a specific example (an exact 
integrable model which has the form of a Toda molecule) and by 
deriving explicit exact solution, we show that 
these models allow for hairy  black brane solutions with non-AdS domain wall 
asymptotics, whose extremal limit is a  scalar soliton.  The soliton smoothly 
interpolates between a non-AdS domain wall 
solution at 
$r=\infty$ and an AdS solution  near $r=0$. 
\end{abstract}


\maketitle



Static black hole and black brane (BB) solutions of Einstein-scalar 
gravity with  non-trivial scalar hair 
and AdS asymptotics are a  crucial ingredient in the recent 
developments of  the AdS/CFT correspondence. 
The coordinate-dependent scalar hair of the black
brane  solutions has a field theory dual interpretation 
as a scalar condensate triggering spontaneous symmetry breaking 
and/or phase transitions \cite{Hartnoll:2008vx,Hartnoll:2008kx,
Horowitz:2008bn,Herzog:2009xv,Hartnoll:2009sz,Charmousis:2009xr,
Cadoni:2009xm,Horowitz:2010gk,Cadoni:2011kv,Gouteraux:2011ce}. Alternatively, the bulk 
scalar can be seen as a running coupling constant and it is very 
useful   for setting up  holographic renormalization group methods 
\cite{Skenderis:2002wp}. 

A particularly simple and interesting example of Einstein-scalar 
gravity is represented by the so-called fake supergravity (SUGRA) 
\cite{DeWolfe:1999cp,Freedman:2003ax,Faulkner:2010fh}. 
In this case, the potential $V(\phi)$ for the scalar field can be derived from a 
superpotential $P(\phi)$ and one can write down first order -- fake 
BPS -- equations, whose solutions automatically satisfy the second-order 
field equations. If there are no singularities, the Witten-Nester 
theorem \cite{Witten:1981mf,Nester:1982tr,Townsend:1984iu}
assures stability of these solitonic solutions.

Until now this scheme has been only  used for potentials $V$ having 
a negative maximum and  with squared mass $m^{2}$ of the corresponding tachyonic
excitation slightly above the Breitenlohner-Freedman (BF) bound, 
$m^{2}_{BF}\le m^{2}<m^{2}_{BF}+1$.  In this range of values of $m^{2}$,
rather generic  boundary conditions for the $r=\infty$ behavior of 
the scalar  field $\phi$ are possible, 
giving rise to the so-called ``designer gravity'' theories, in which 
the mass of the solitons can be pre-ordered \cite{Hertog:2004ns}.

In the case of spherical solutions,  boundary conditions can be 
found, which break the AdS symmetries and allow for stable solitonic 
solutions \cite{Battarra:2011nt}.
But, unfortunately, in the case of planar solutions only AdS-symmetry 
preserving boundary conditions are possible. These boundary conditions 
do not allow for a stable ground state unless the potential $V$ has a 
second extremum \cite{Faulkner:2010fh}. If this is not the case, the scalar behaves 
logarithmically and the  solution  interpolates between AdS at 
$r=\infty$ and a domain wall (DW) near $r=0$, which is singular at $r=0$. 

In this letter we will consider potentials $V$ that have a minimum 
instead of a maximum and standard, AdS-symmetry preserving, Dirichlet 
boundary conditions for the $r=\infty$ behavior of the scalar $\phi$.
In this case, positive-energy theorems (PET) allow for a stable 
ground state solitonic solution, but standard no-hair theorems forbid the 
existence of BB solutions with a regular horizon 
\cite{Torii:2001pg,Hertog:2006rr}.
For this reason models whose  potential $V$ has a minimum  have not  been 
taken into consideration in this context.

However, recently formulated  no-hair theorems indicate that only BB solutions with 
AdS asymptotics are forbidden \cite{Cadoni:2011nq}, leaving open the possibility of having 
BB solutions with generic domain wall asymptotics. We show that 
this is the case by considering  a specific example. We  will 
consider a fake SUGRA model, which can be recast in the form of a 
Toda molecule and  is exactly integrable.
We derive explicit exact solutions and show that 
the model allows for black brane solutions with non-AdS domain wall 
asymptotics, whose extremal limit is a scalar soliton. This 
 soliton smoothly interpolates between a non-AdS domain wall solution at 
$r=\infty$ and an AdS solution  near $r=0$.

Let us investigate static, radially symmetric, planar solutions of Einstein gravity
 minimally coupled  to a scalar field with self-interaction potential $V(\phi)$. 
 The  action is
\beq
S=\int d^{4}x \sqrt{-g}\left[R-2 (\partial \phi)^{2} -V(\phi)\right].\label{lagr}
\eeq
We assume that $V(\phi)$ has a negative minimum at $\phi=0$, thus 
allowing an AdS$_{4}$ vacuum,  corresponding to a 
positive squared mass $m^{2}$ for the scalar excitation. For $\phi$, we    
adopt standard (Dirichlet) boundary conditions at $r=\infty$, which preserve the 
asymptotic symmetry group of AdS$_{4}$.   
We also assume that 
$V(\phi)$ can be derived from a superpotential $P(\phi)$,
\beq\lb{k1}
V(\phi)= 2\left(\frac{dP}{d\phi}\right)^{2}-6P^{2}.
\eeq
This means that our theory is a fake SUGRA  model 
\cite{DeWolfe:1999cp,Freedman:2003ax,Faulkner:2010fh}, namely one 
can define a spinor energy using fake transformations similar to real 
SUGRA theories. In particular, if we 
parametrize the spacetime metric as $ds^{2}= 
r^{2}(-dt^{2}+dx_{i}dx^{i})+ h^{-1}dr^{2}$, the second-order field 
equations stemming from (\ref{lagr}) reduce to first order equations 
\cite{DeWolfe:1999cp,Freedman:2003ax,Faulkner:2010fh}
\beq\lb{k2}
\phi'(r)=- \frac{P_{,\phi}¥}{rP(\phi)},\quad 
h(r)=r^{2}P^{2}(\phi).
\eeq
Using  the standard Witten-Nester procedure 
\cite{Witten:1981mf,Nester:1982tr,Townsend:1984iu} one can then  show 
that the energy of any singularity-free solution of the  first-order 
equations (\ref{k2}) is bounded from below.

For definiteness, we will focus on a fake SUGRA model defined by ($L$ 
is the AdS length)
\beq\lb{k3}
V(\phi)= -\frac{6}{\g L^{2}}\left(e^{2\sqrt3\beta 
\phi}-\beta^{2}e^{\frac{2\sqrt3}{\beta}\phi}\right), \quad P(\phi)=
\frac{1}{\g L}\left(e^{\sqrt3\beta 
\phi}-\beta^{2}e^{\frac{\sqrt3}{\beta}\phi}\right),\quad 
\g=1-\b^2.\eeq
The potential is defined for  every $\beta\neq 0,1$.
It  has always a  minimum at $\phi=0$, with $V(0)=-6/L^{2}$, 
corresponding to the AdS$_{4}$ solution and to a scalar 
excitation with positive squared mass  $m^{2}=18/L^{2}$.
We use standard (Dirichlet) 
boundary conditions  for $\phi$, which set to zero the dominant term 
in the $r\to\infty$ expansion.  The   fall-off behavior
of the scalar field    is  therefore given by
$\phi\sim \frac{\beta}{r^{6}}.$

The above-mentioned stability theorem allows in principle  for the 
existence of a stable 
ground state  hairy solitonic solution, but standard no-hair theorems forbid the 
existence of BB solutions with AdS asymptotics when $m^{2}$ is 
positive \cite{Torii:2001pg,Hertog:2006rr}. Even if a solitonic solution exists, it cannot 
be obtained as the extremal limit of an asymptotically AdS solution. 
We will therefore look for BB solutions of 
(\ref{lagr}) with    asymptotics, 
\beq\lb{DW}
ds^{2}= r^{\eta}(-dt^{2}+ 
dx_{i}dx^{i}) + r^{-\eta}dr^{2},
\eeq
with $0\le \eta\le 2$. For 
$\eta=0,2$, Eq. (\ref{DW}) describes flat or AdS spacetime, respectively.
When $0<\eta<2$ (\ref{DW}) describes a brane, which we  call non-AdS 
domain wall. 
This kind of spacetimes have been already investigated in the 
literature. In particular, it has been show that they admit an 
holographic interpretation for $1\le \eta\le 2$ 
\cite{Kaitscheider:2009as,Boonstra:1998mp}.

The field equations of the Einstein-scalar gravity model with potential 
(\ref{k3}) can be exactly  integrated. This can be achieved using a 
parametrization of the metric introduced in \cite{Gibbons:1987ps} 
and used in several investigations 
of dilatonic black holes 
\cite{Mignemi:1988qc,Wiltshire:1990ah,Poletti:1994ff,Mignemi:1999zy,Mignemi:2006ut,
Cadoni:1993yt,Monni:1995vu}
\beq\lb{met}
ds^2=-\en dt^2+e^{2\n+4\r}d\x^2+\er dx_{i}dx^{i}.
\eeq

Using this parametrization,  the field equations can be 
recast in the form of the $SU(2)\times SU(2)$ Toda molecule \cite{Olshanetsky:1981dk}.
In fact, defining new variables $\O= \nu+ 2 \r +{\sqrt3\b} \phi,\, 
\S= \nu+ 2 \r +\frac{\sqrt3}{\b}\phi,$ and taking into account that 
the field equations imply  $\r=\n+c\x$, with $c$ an integration 
constant, one obtains the second-order equations
\beq\lb{fe25}
\ddot\O=\frac{9}{L^{2}}e^{2\O},\qquad\ddot\S=\frac{9}{L^{2}}e^{2\S},
\eeq
subject to the constraint 
\beq\lb{fe26}
\dot\O^2-\b^2\dot\S^2-\g c^2=\frac{9}{L^{2}}(e^{2\O}-\b^2e^{2\S}).
\eeq
These equations can  be solved to give the general solution
\bea\lb{solution}
\en&=&\left(\frac{2L}{3}\right)^{2/3} a^{\frac{2}{3\g}} b^{\frac{-2\b^{2}}{3\g}}
e^{\frac{2b\beta^{2}\x_{0}}{3\g}}e^{2(a-\b^2b-2\g c)\x/3\g}
\left[{(1-e^{2b(\x-\x_0)})^{\b^2}
\over1-\eax}\right]^{2/3\g},\cr
\er&=&\left(\frac{2L}{3}\right)^{2/3} a^{\frac{2}{3\g}} b^{\frac{-2\b^{2}}{3\g}}
e^{\frac{2b\beta^{2}\x_{0}}{3\g}}e^{2(a-\b^2b+\g c)\x/3\g}
\left[{(1-e^{2b(\x-\x_0)})^{\b^2}
\over1-\eax}\right]^{2/3\g},\cr
\f&=&\frac{\b}{\sqrt3\,\g}\log\left[\frac{b\sinh a\x}{a\sinh b(\x-\x_0)}\right],
\eea
where $\x_{0}$ is an arbitrary integration constant and $a,b,c$ must satisfy 
the constraint $\g c^{2}=a^{2}-\b^2 b^{2}$.

We are interested in solutions with a regular horizon at $\x=\x_h$.
Requiring $e^{2\nu}(\xi_{h})=0$ and $e^{2\r}(\xi_{h})=$ const, 
one easily realizes that this is only possible for
$\xi_{h}\to-\infty$, when $\g c=\beta^2b-a$. This condition, together with the 
constraint, implies $a=b=-c$. In the case $\x_0=0$, we obtain
the planar Schwarzschild-anti de Sitter solution with $\phi=0$. As 
one can show by expanding (\ref{solution}) near $\xi=0$ and 
$\xi=-\infty$,  all 
the other 
solutions with AdS asymptotics and non-trivial scalar hair have a 
naked singularity at $r=0$ with $\phi\sim \log r$. 
This is in  complete accordance with the results of well-established  
no-hair theorems. 

In the general case $\x_0\ne0$ we have  solutions with a regular 
horizon, but they do not approach  AdS$_{4}$ asymptotically,
and it is not possible to write them  in a \sch form
in terms of elementary functions.
Let us first consider the case $\b^{2}<1$.  
In this case the asymptotic region  corresponds to the limit $\x\to0$.
Defining the  new radial coordinate  
$\sigma r=(1-\eax)^{-(1+3\b^2)/3\g}$ with $\sigma$ constant,  for 
$0<\x_{0}<\infty$ the solution (\ref{solution}) becomes, 
\bea\lb{met1}
ds^2=&&\left(1+\frac{\m_2}{r^\d}\right)^{2\b^2/3\g}\left[-\left(1-\frac{\m_1}
{r^\d}\right)r^{2/(1+3\b^2)}dt^2+
\frac{E(1+\m_2/r^{\d})^{4\b^2/3\g}\,dr^2}{(1-\m_1/r^{\d}) r^{2/(1+3\b^2)}}
+ r^{2/(1+3\b^2)}\OM\right],\cr
e^{2\f}=&&D\left(1+\frac{\m_2}{r^\d}\right)^{-2\b/\sqrt3\,\g}r^{-2\sqrt3\,\b/(1+3\b^2)},
\eea
where $\m_1\ge 0,\m_{2}>0$ are  free parameters, $\d=3\g/(1+3\b^{2})\,$, 
$D= [\m_{2}(\m_{1}+\m_{2})]^{\b/\sqrt3\g}$, and  
$E=[\g L/(1+3\b^{2}¥)]^{2}¥D^{-\sqrt3 \b}¥$.

The  asymptotic behavior of this solution for $r\to\infty$ is 
that of a domain wall (\ref{DW}) with $\eta=2/(1+3\b^2)$ and $\phi= 
-[(\sqrt{3}\b)/(1+3\b^{2})]\ln r$.  
For $\m_1>0$, the metric (\ref{met1}) exhibits a singularity at $r=0$ 
shielded by a horizon at $r=\m_1^{\,1/\d}$, and therefore represents a 
regular black brane. Owing to the fact that the scalar $\phi$ 
depends on $\m_{1}$,  the  existence of this BB solution is perfectly 
consistent with the no-hair theorem of Ref. \cite{Cadoni:2011nq}.
Notice that although the scalar field  remains finite at $r=0$, 
the scalar curvature $R$ of spacetime diverges as $ 
R\sim r^{-3(1+\b^{2})(1+3\b^{2})}$.
The extremal, zero temperature, solution is obtained for $\m_{1}=0$,
\beq\lb{f5}
ds^2=\left(1+\frac{\m_2}{r^\d}\right)^{2\b^2/3\g}\left[
r^{2/(1+3\b^2)}\left(-dt^2+\OM\right)+
E r^{-2/(1+3\b^2)}(1+\frac{\m_2}{r^{\d}})^{4\b^2/3\g}\,dr^2\right],
\eeq
while the scalar field maintains the form of Eq.\ (\ref{met1}).
The extremal solution (\ref{f5})  represents a regular soliton. In 
fact, not only the scalar field is finite at $r=0$ 
($e^{2\phi}=D(\m_{2})^{-(2\b)/(\sqrt{3}\g)}$) 
but also the scalar curvature of the spacetime   remains finite both 
ar $r=0$ and $r=\infty$. 
The extremal soliton has the form of a  brane, for which the 
metric behaves for small and large $r$ as in Eq. (\ref{DW})
with a different power of $r$ in the $r=\infty$ and $r=0$ 
region. Whereas for $r\to\infty$, we have  $\eta=2/(1+3\b^2)$ and 
$\phi\sim \ln r$, near the origin 
$\eta= 2$ and $\phi=const.$. Hence, our soliton (\ref{f5}) 
interpolates between a DW solution at infinity  and AdS 
spacetime at $r=0$.
As expected the soliton (\ref{f5}) satisfies the fake BPS equations 
(\ref{k2}).

A similar procedure allows one to find the solution when $\b^2>1$. 
Now the asymptotic region $r\to\inf$ corresponds  $\x\to\x_0$. 
As before, the metric can be written in terms of a new radial coordinate 
$\sigma r=(1-e^{2a(\x-\x_{0})}¥)^{(3+\b^2)/3\g}$, 
\bea\lb{met2}
ds^2&&=\left(1+\frac{\m_2}{r^\d}\right)^{-2/3\g}\left[-\left(1-\frac{\m_1}{r^\d}
\right)r^{2\b^2/(3+\b^2)}dt^2+\frac{E(1+\m_2/r^\d)^{-4/3\g}dr^2}
{(1-\m_1/r^\d)r^{2\b^2/(3+\b^2)}}+r^{2\b^2/(3+\b^2)}\OM\ \right],\cr
e^{2\f}&&= D\left(1+\frac{\m_2}{r^\d}\right)^{2\b/\sqrt3\,\g}r^{-2\sqrt3\,\b/(3+\b^2)},
\eea
where now $\d=-3\g/(3+\b^2)>0$,  
$D= [\m_{2}(\m_{1}+\m_{2})]^{\b/\sqrt3\g}$, and  
$E=[\g L/(3+\b^{2})]^{2}D^{-\sqrt3 \b}¥$.
At infinity the solution behaves as a domain wall with 
$\eta=2\b^2/(3+\b^2)$  and $\phi=-[(\sqrt{3}\b)/(3+\b^{2})]\ln r$.

As in the previous case, if $\m_{1}>0$, the metric 
exhibits a singularity at $r=0 $ and a
horizon at $r=\m_1^{\,1/\d}$, and therefore describes a 
regular black brane with non-AdS domain wall asymptotics.

Also in this case the extremal, zero temperature solution, obtained 
for $\m_{1}=0$, is a regular soliton that satisfies Eq.\ (\ref{k2}),
\beq\lb{f9}
ds^2=\left(1+\frac{\m_2}{r^\d}\right)^{-2/3\g}\left[
r^{2\b^{2}¥/(3+\b^2)}\left(-dt^2+\OM\right)+
E r^{-2\b^{2}¥/(3+\b^2)}(1+\frac{\m_2}{r^{\d}})^{-4/3\g}\,dr^2\right].
\eeq
As expected, the soliton interpolates between the domain wall solution 
(\ref{DW}) with $\eta=2\b^2/(3+\b^2)$ at infinity and an AdS solution with 
constant $\phi$ near $r=0$.

It may be interesting to notice that the Schwarzschild-anti de Sitter solution
is recovered in the singular limit $\m_2\to\inf$ of (\ref{met1}) or (\ref{met2}).

The question about the stability of our scalar solitonic solutions 
(\ref{f5}),  (\ref{f9}) is rather involved. Stability cannot be simply shown 
using the standard Witten-Nester procedure. In fact, the standard 
demonstration requires an asymptotic AdS (or flat) spacetime, whereas 
our solutions have non-AdS DW asymptotics.  This issue will be 
investigated in a forthcoming paper. 

Let us now compare our results with those 
obtained when the potential has a negative maximum  
with $m^{2}_{BF}\le m^{2}<m^{2}_{BF}+1$.
If the potential $V(\phi)$ behaves exponentially at large $\phi$, one has 
 solutions with  AdS$_{4}$ asymptotics at large $r$ and 
singular DW behavior near $r=0$, with $\phi\sim \ln r$ 
\cite{Faulkner:2010fh,
Cadoni:2011nq}.
 The only known case that
does not present a small-$r$ singularity is when $V$ has a second 
extremum. Apart from this case,  the solutions always  have  opposite 
behavior with respect  to the soliton that we get in the  $m^{2}>0$ case: 
the solution interpolates between an AdS$_{4}$ spacetime at $r=\infty$ 
and  a DW solution near $r=0$ \cite{Cadoni:2011nq}.  

In this context, it is also interesting to notice that also a pure 
exponential potential $V=-2\l e^{-2h\phi}$ for $h^{2}<3$ is a  fake 
SUGRA model \cite{Faulkner:2010fh}. In fact, $V$ can be derived from the 
superpotential $P=\sqrt{\l/(3-h^{2})}\,e^{-h \phi}$.
Also in this case the field equation can be exactly 
integrated using the Toda molecule parametrization (\ref{met}) for 
the metric. BB solutions with DW asymptotics  can be found using the 
procedure described above. Defining  a new variable 
$\y=\n+2\r-h\f$, the field equations can be recast in the form
$\ddot\y=(3-h^2)\l\ee$, together with a constraint involving
the integration constants. Solving these equations, one can show that 
the solutions 
with a regular horizon can be written in the form 
$ds^{2}=-U(r) dt^{2}+U(r)^{-1}dr^{2}+ R(r)^{2}\OM,$
with
$$ U=\left(1-\m r^{(h^2-3)/(1+h^2)}\right)r^{2/(1+h^2)},\quad
R(r)=r^{1/(1+h^2)},\quad\ex^{2\f}=C r^{2h/(1+h^2)},$$
where  $\m$ is an integration constant and $C=[(\l(1+h^2)^2/(2(3-h^{2}¥))]^{1/h}$. 
For $\m=0$ we get 
a  DW solution, which is singular at $r=0$. This form of the solution 
has been already derived in Ref.\ \cite{Cadoni:2011nq}, 
using  a different method. 

In this letter we have derived explicit exact black brane solutions 
of Einstein-scalar gravity with positive  squared mass for the scalar 
field,  whose 
extremal limit is a regular scalar soliton. We have circumvented  
standard no-hair theorems by allowing for solutions with non-AdS 
domain wall asymptotics. We have derived the solutions for 4D 
Einstein-scalar gravity but our derivation could be easily extended 
to arbitrary spacetime dimensions.  The scalar soliton interpolates 
between AdS$_{4}$ for small $r$ and non-AdS brane at large $r$. The 
soliton  has an 
holographic interpretation in terms of a flow of a dual 3D QFT 
between n IR 
fixed point at $r=0$ and an UV Poincar\'e invariant vacuum at 
$r=\infty$. Hence, our results  may have very useful  
applications in the  AdS/CFT correspondence context.

\bibliography{dw}

\begin{thebibliography}{33}
\expandafter\ifx\csname natexlab\endcsname\relax\def\natexlab#1{#1}\fi
\expandafter\ifx\csname bibnamefont\endcsname\relax
  \def\bibnamefont#1{#1}\fi
\expandafter\ifx\csname bibfnamefont\endcsname\relax
  \def\bibfnamefont#1{#1}\fi
\expandafter\ifx\csname citenamefont\endcsname\relax
  \def\citenamefont#1{#1}\fi
\expandafter\ifx\csname url\endcsname\relax
  \def\url#1{\texttt{#1}}\fi
\expandafter\ifx\csname urlprefix\endcsname\relax\def\urlprefix{URL }\fi
\providecommand{\bibinfo}[2]{#2}
\providecommand{\eprint}[2][]{\url{#2}}

\bibitem[{\citenamefont{Hartnoll
  et~al.}(2008{\natexlab{a}})\citenamefont{Hartnoll, Herzog, and
  Horowitz}}]{Hartnoll:2008vx}
\bibinfo{author}{\bibfnamefont{S.~A.} \bibnamefont{Hartnoll}},
  \bibinfo{author}{\bibfnamefont{C.~P.} \bibnamefont{Herzog}},
  \bibnamefont{and} \bibinfo{author}{\bibfnamefont{G.~T.}
  \bibnamefont{Horowitz}}, \bibinfo{journal}{Phys. Rev. Lett.}
  \textbf{\bibinfo{volume}{101}}, \bibinfo{pages}{031601}
  (\bibinfo{year}{2008}{\natexlab{a}}), \eprint{0803.3295}.

\bibitem[{\citenamefont{Hartnoll
  et~al.}(2008{\natexlab{b}})\citenamefont{Hartnoll, Herzog, and
  Horowitz}}]{Hartnoll:2008kx}
\bibinfo{author}{\bibfnamefont{S.~A.} \bibnamefont{Hartnoll}},
  \bibinfo{author}{\bibfnamefont{C.~P.} \bibnamefont{Herzog}},
  \bibnamefont{and} \bibinfo{author}{\bibfnamefont{G.~T.}
  \bibnamefont{Horowitz}}, \bibinfo{journal}{JHEP}
  \textbf{\bibinfo{volume}{12}}, \bibinfo{pages}{015}
  (\bibinfo{year}{2008}{\natexlab{b}}), \eprint{0810.1563}.

\bibitem[{\citenamefont{Horowitz and Roberts}(2008)}]{Horowitz:2008bn}
\bibinfo{author}{\bibfnamefont{G.~T.} \bibnamefont{Horowitz}} \bibnamefont{and}
  \bibinfo{author}{\bibfnamefont{M.~M.} \bibnamefont{Roberts}},
  \bibinfo{journal}{Phys.Rev.} \textbf{\bibinfo{volume}{D78}},
  \bibinfo{pages}{126008} (\bibinfo{year}{2008}), \eprint{0810.1077}.

\bibitem[{\citenamefont{Herzog}(2009)}]{Herzog:2009xv}
\bibinfo{author}{\bibfnamefont{C.~P.} \bibnamefont{Herzog}},
  \bibinfo{journal}{J. Phys.} \textbf{\bibinfo{volume}{A42}},
  \bibinfo{pages}{343001} (\bibinfo{year}{2009}), \eprint{0904.1975}.

\bibitem[{\citenamefont{Hartnoll}(2009)}]{Hartnoll:2009sz}
\bibinfo{author}{\bibfnamefont{S.~A.} \bibnamefont{Hartnoll}}
  (\bibinfo{year}{2009}), \eprint{0903.3246}.

\bibitem[{\citenamefont{Charmousis et~al.}(2009)\citenamefont{Charmousis,
  Gouteraux, and Soda}}]{Charmousis:2009xr}
\bibinfo{author}{\bibfnamefont{C.}~\bibnamefont{Charmousis}},
  \bibinfo{author}{\bibfnamefont{B.}~\bibnamefont{Gouteraux}},
  \bibnamefont{and} \bibinfo{author}{\bibfnamefont{J.}~\bibnamefont{Soda}},
  \bibinfo{journal}{Phys. Rev.} \textbf{\bibinfo{volume}{D80}},
  \bibinfo{pages}{024028} (\bibinfo{year}{2009}), \eprint{0905.3337}.

\bibitem[{\citenamefont{Cadoni et~al.}(2010)\citenamefont{Cadoni, D'Appollonio,
  and Pani}}]{Cadoni:2009xm}
\bibinfo{author}{\bibfnamefont{M.}~\bibnamefont{Cadoni}},
  \bibinfo{author}{\bibfnamefont{G.}~\bibnamefont{D'Appollonio}},
  \bibnamefont{and} \bibinfo{author}{\bibfnamefont{P.}~\bibnamefont{Pani}},
  \bibinfo{journal}{JHEP} \textbf{\bibinfo{volume}{03}}, \bibinfo{pages}{100}
  (\bibinfo{year}{2010}), \eprint{0912.3520}.

\bibitem[{\citenamefont{Horowitz}(2010)}]{Horowitz:2010gk}
\bibinfo{author}{\bibfnamefont{G.~T.} \bibnamefont{Horowitz}}
  (\bibinfo{year}{2010}), \eprint{1002.1722}.

\bibitem[{\citenamefont{Cadoni and Pani}(2011)}]{Cadoni:2011kv}
\bibinfo{author}{\bibfnamefont{M.}~\bibnamefont{Cadoni}} \bibnamefont{and}
  \bibinfo{author}{\bibfnamefont{P.}~\bibnamefont{Pani}},
  \bibinfo{journal}{JHEP} \textbf{\bibinfo{volume}{1104}}, \bibinfo{pages}{049}
  (\bibinfo{year}{2011}), \eprint{1102.3820}.

\bibitem[{\citenamefont{Gouteraux and Kiritsis}(2011)}]{Gouteraux:2011ce}
\bibinfo{author}{\bibfnamefont{B.}~\bibnamefont{Gouteraux}} \bibnamefont{and}
  \bibinfo{author}{\bibfnamefont{E.}~\bibnamefont{Kiritsis}}
  (\bibinfo{year}{2011}), \eprint{1107.2116}.

\bibitem[{\citenamefont{Skenderis}(2002)}]{Skenderis:2002wp}
\bibinfo{author}{\bibfnamefont{K.}~\bibnamefont{Skenderis}},
  \bibinfo{journal}{Class.Quant.Grav.} \textbf{\bibinfo{volume}{19}},
  \bibinfo{pages}{5849} (\bibinfo{year}{2002}), \eprint{hep-th/0209067}.

\bibitem[{\citenamefont{DeWolfe et~al.}(2000)\citenamefont{DeWolfe, Freedman,
  Gubser, and Karch}}]{DeWolfe:1999cp}
\bibinfo{author}{\bibfnamefont{O.}~\bibnamefont{DeWolfe}},
  \bibinfo{author}{\bibfnamefont{D.}~\bibnamefont{Freedman}},
  \bibinfo{author}{\bibfnamefont{S.}~\bibnamefont{Gubser}}, \bibnamefont{and}
  \bibinfo{author}{\bibfnamefont{A.}~\bibnamefont{Karch}},
  \bibinfo{journal}{Phys.Rev.} \textbf{\bibinfo{volume}{D62}},
  \bibinfo{pages}{046008} (\bibinfo{year}{2000}), \eprint{hep-th/9909134}.

\bibitem[{\citenamefont{Freedman et~al.}(2004)\citenamefont{Freedman, Nunez,
  Schnabl, and Skenderis}}]{Freedman:2003ax}
\bibinfo{author}{\bibfnamefont{D.}~\bibnamefont{Freedman}},
  \bibinfo{author}{\bibfnamefont{C.}~\bibnamefont{Nunez}},
  \bibinfo{author}{\bibfnamefont{M.}~\bibnamefont{Schnabl}}, \bibnamefont{and}
  \bibinfo{author}{\bibfnamefont{K.}~\bibnamefont{Skenderis}},
  \bibinfo{journal}{Phys.Rev.} \textbf{\bibinfo{volume}{D69}},
  \bibinfo{pages}{104027} (\bibinfo{year}{2004}), \eprint{hep-th/0312055}.

\bibitem[{\citenamefont{Faulkner et~al.}(2010)\citenamefont{Faulkner, Horowitz,
  and Roberts}}]{Faulkner:2010fh}
\bibinfo{author}{\bibfnamefont{T.}~\bibnamefont{Faulkner}},
  \bibinfo{author}{\bibfnamefont{G.~T.} \bibnamefont{Horowitz}},
  \bibnamefont{and} \bibinfo{author}{\bibfnamefont{M.~M.}
  \bibnamefont{Roberts}}, \bibinfo{journal}{Class.Quant.Grav.}
  \textbf{\bibinfo{volume}{27}}, \bibinfo{pages}{205007}
  (\bibinfo{year}{2010}), \eprint{1006.2387}.

\bibitem[{\citenamefont{Witten}(1981)}]{Witten:1981mf}
\bibinfo{author}{\bibfnamefont{E.}~\bibnamefont{Witten}},
  \bibinfo{journal}{Commun.Math.Phys.} \textbf{\bibinfo{volume}{80}},
  \bibinfo{pages}{381} (\bibinfo{year}{1981}).

\bibitem[{\citenamefont{Nester}(1981)}]{Nester:1982tr}
\bibinfo{author}{\bibfnamefont{J.~A.} \bibnamefont{Nester}},
  \bibinfo{journal}{Phys.Lett.} \textbf{\bibinfo{volume}{A83}},
  \bibinfo{pages}{241} (\bibinfo{year}{1981}).

\bibitem[{\citenamefont{Townsend}(1984)}]{Townsend:1984iu}
\bibinfo{author}{\bibfnamefont{P.}~\bibnamefont{Townsend}},
  \bibinfo{journal}{Phys.Lett.} \textbf{\bibinfo{volume}{B148}},
  \bibinfo{pages}{55} (\bibinfo{year}{1984}).

\bibitem[{\citenamefont{Hertog and Horowitz}(2005)}]{Hertog:2004ns}
\bibinfo{author}{\bibfnamefont{T.}~\bibnamefont{Hertog}} \bibnamefont{and}
  \bibinfo{author}{\bibfnamefont{G.~T.} \bibnamefont{Horowitz}},
  \bibinfo{journal}{Phys.Rev.Lett.} \textbf{\bibinfo{volume}{94}},
  \bibinfo{pages}{221301} (\bibinfo{year}{2005}), \eprint{hep-th/0412169}.

\bibitem[{\citenamefont{Battarra}(2011)}]{Battarra:2011nt}
\bibinfo{author}{\bibfnamefont{L.}~\bibnamefont{Battarra}}
  (\bibinfo{year}{2011}), \eprint{1110.1083}.

\bibitem[{\citenamefont{Torii et~al.}(2001)\citenamefont{Torii, Maeda, and
  Narita}}]{Torii:2001pg}
\bibinfo{author}{\bibfnamefont{T.}~\bibnamefont{Torii}},
  \bibinfo{author}{\bibfnamefont{K.}~\bibnamefont{Maeda}}, \bibnamefont{and}
  \bibinfo{author}{\bibfnamefont{M.}~\bibnamefont{Narita}},
  \bibinfo{journal}{Phys.Rev.} \textbf{\bibinfo{volume}{D64}},
  \bibinfo{pages}{044007} (\bibinfo{year}{2001}).

\bibitem[{\citenamefont{Hertog}(2006)}]{Hertog:2006rr}
\bibinfo{author}{\bibfnamefont{T.}~\bibnamefont{Hertog}},
  \bibinfo{journal}{Phys. Rev.} \textbf{\bibinfo{volume}{D74}},
  \bibinfo{pages}{084008} (\bibinfo{year}{2006}), \eprint{gr-qc/0608075}.

\bibitem[{\citenamefont{Cadoni et~al.}(2011)\citenamefont{Cadoni, Mignemi, and
  Serra}}]{Cadoni:2011nq}
\bibinfo{author}{\bibfnamefont{M.}~\bibnamefont{Cadoni}},
  \bibinfo{author}{\bibfnamefont{S.}~\bibnamefont{Mignemi}}, \bibnamefont{and}
  \bibinfo{author}{\bibfnamefont{M.}~\bibnamefont{Serra}},
  \bibinfo{journal}{Phys.Rev.} \textbf{\bibinfo{volume}{D84}},
  \bibinfo{pages}{084046} (\bibinfo{year}{2011}), \eprint{1107.5979}.

\bibitem[{\citenamefont{Kanitscheider and
  Skenderis}(2009)}]{Kaitscheider:2009as}
\bibinfo{author}{\bibfnamefont{I.}~\bibnamefont{Kanitscheider}}
  \bibnamefont{and}
  \bibinfo{author}{\bibfnamefont{K.}~\bibnamefont{Skenderis}},
  \bibinfo{journal}{JHEP} \textbf{\bibinfo{volume}{0904}}, \bibinfo{pages}{062}
  (\bibinfo{year}{2009}), \eprint{0901.1487}.

\bibitem[{\citenamefont{Boonstra et~al.}(1999)\citenamefont{Boonstra,
  Skenderis, and Townsend}}]{Boonstra:1998mp}
\bibinfo{author}{\bibfnamefont{H.}~\bibnamefont{Boonstra}},
  \bibinfo{author}{\bibfnamefont{K.}~\bibnamefont{Skenderis}},
  \bibnamefont{and} \bibinfo{author}{\bibfnamefont{P.}~\bibnamefont{Townsend}},
  \bibinfo{journal}{JHEP} \textbf{\bibinfo{volume}{9901}}, \bibinfo{pages}{003}
  (\bibinfo{year}{1999}), \eprint{hep-th/9807137}.

\bibitem[{\citenamefont{Gibbons and Maeda}(1988)}]{Gibbons:1987ps}
\bibinfo{author}{\bibfnamefont{G.}~\bibnamefont{Gibbons}} \bibnamefont{and}
  \bibinfo{author}{\bibfnamefont{K.-i.} \bibnamefont{Maeda}},
  \bibinfo{journal}{Nucl.Phys.} \textbf{\bibinfo{volume}{B298}},
  \bibinfo{pages}{741} (\bibinfo{year}{1988}).

\bibitem[{\citenamefont{Mignemi and Wiltshire}(1989)}]{Mignemi:1988qc}
\bibinfo{author}{\bibfnamefont{S.}~\bibnamefont{Mignemi}} \bibnamefont{and}
  \bibinfo{author}{\bibfnamefont{D.}~\bibnamefont{Wiltshire}},
  \bibinfo{journal}{Class.Quant.Grav.} \textbf{\bibinfo{volume}{6}},
  \bibinfo{pages}{987} (\bibinfo{year}{1989}).

\bibitem[{\citenamefont{Wiltshire}(1991)}]{Wiltshire:1990ah}
\bibinfo{author}{\bibfnamefont{D.~L.} \bibnamefont{Wiltshire}},
  \bibinfo{journal}{Phys.Rev.} \textbf{\bibinfo{volume}{D44}},
  \bibinfo{pages}{1100} (\bibinfo{year}{1991}).

\bibitem[{\citenamefont{Poletti and Wiltshire}(1994)}]{Poletti:1994ff}
\bibinfo{author}{\bibfnamefont{S.}~\bibnamefont{Poletti}} \bibnamefont{and}
  \bibinfo{author}{\bibfnamefont{D.}~\bibnamefont{Wiltshire}},
  \bibinfo{journal}{Phys.Rev.} \textbf{\bibinfo{volume}{D50}},
  \bibinfo{pages}{7260} (\bibinfo{year}{1994}), \eprint{gr-qc/9407021}.

\bibitem[{\citenamefont{Mignemi}(2000)}]{Mignemi:1999zy}
\bibinfo{author}{\bibfnamefont{S.}~\bibnamefont{Mignemi}},
  \bibinfo{journal}{Phys.Rev.} \textbf{\bibinfo{volume}{D62}},
  \bibinfo{pages}{024014} (\bibinfo{year}{2000}), \eprint{gr-qc/9910041}.

\bibitem[{\citenamefont{Mignemi}(2006)}]{Mignemi:2006ut}
\bibinfo{author}{\bibfnamefont{S.}~\bibnamefont{Mignemi}},
  \bibinfo{journal}{Phys.Rev.} \textbf{\bibinfo{volume}{D74}},
  \bibinfo{pages}{124008} (\bibinfo{year}{2006}), \eprint{gr-qc/0607005}.

\bibitem[{\citenamefont{Cadoni and Mignemi}(1993)}]{Cadoni:1993yt}
\bibinfo{author}{\bibfnamefont{M.}~\bibnamefont{Cadoni}} \bibnamefont{and}
  \bibinfo{author}{\bibfnamefont{S.}~\bibnamefont{Mignemi}},
  \bibinfo{journal}{Phys.Rev.} \textbf{\bibinfo{volume}{D48}},
  \bibinfo{pages}{5536} (\bibinfo{year}{1993}), \eprint{hep-th/9305107}.

\bibitem[{\citenamefont{Monni and Cadoni}(1996)}]{Monni:1995vu}
\bibinfo{author}{\bibfnamefont{S.}~\bibnamefont{Monni}} \bibnamefont{and}
  \bibinfo{author}{\bibfnamefont{M.}~\bibnamefont{Cadoni}},
  \bibinfo{journal}{Nucl. Phys.} \textbf{\bibinfo{volume}{B466}},
  \bibinfo{pages}{101} (\bibinfo{year}{1996}), \eprint{hep-th/9511067}.

\bibitem[{\citenamefont{Olshanetsky and Perelomov}(1981)}]{Olshanetsky:1981dk}
\bibinfo{author}{\bibfnamefont{M.}~\bibnamefont{Olshanetsky}} \bibnamefont{and}
  \bibinfo{author}{\bibfnamefont{A.}~\bibnamefont{Perelomov}},
  \bibinfo{journal}{Phys.Rept.} \textbf{\bibinfo{volume}{71}},
  \bibinfo{pages}{313} (\bibinfo{year}{1981}).

\end{thebibliography}
\end{document}